\documentstyle[11pt]{article}
\def\title#1{\vskip 0.2cm \begin{center}{\Large\bf #1}\end{center}}
\def\authors#1{\begin{center}{\bf #1}\end{center}}
\def\affiliation#1{\begin{center}{\small\it #1}\end{center}}
\def\abstract#1{\vskip 0.5cm {#1}}
\def\msol{{\cal M}_\odot}
\def\lsol{{\cal L}_{\odot,B}}
\def\lso{{\cal L}_{\odot}}
\def \refs {\begingroup \frenchspacing
            \parindent = 0 pt
            \everypar = {\hangindent = 20.0 pt \hangafter = 1}}
\def \endrefs {\par \endgroup}
\input psfig.tex
\begin{document}

\title{Cluster-Group Interaction in the Virgo Cluster}

\authors{S. Schindler$^{}$}

\affiliation{
    University of Innsbruck, Institute for Astrophysics,
        Technikerstr. 25, 6020 Innsbruck, Austria}

\abstract{We present two projects related to interaction in the Virgo
cluster. In the first section we draw a quantitative comparison of the
distribution of the galaxies and the intra-cluster gas taking into
account that the Virgo cluster has an irregular structure consisting
of several subclusters. In 
the second section we show hydrodynamic simulations of the interaction
(ram pressure stripping) 
of a galaxy like M86 with the intra-cluster gas.}

\section{Comparison of the optical and the X-ray morphology of the Virgo cluster}

The Virgo cluster, as the nearest cluster of galaxies, is ideally suited
for a detailed comparison of the distribution of two
components -- intra-cluster gas and galaxies. 

\begin{figure}[ht]
\psfig{figure=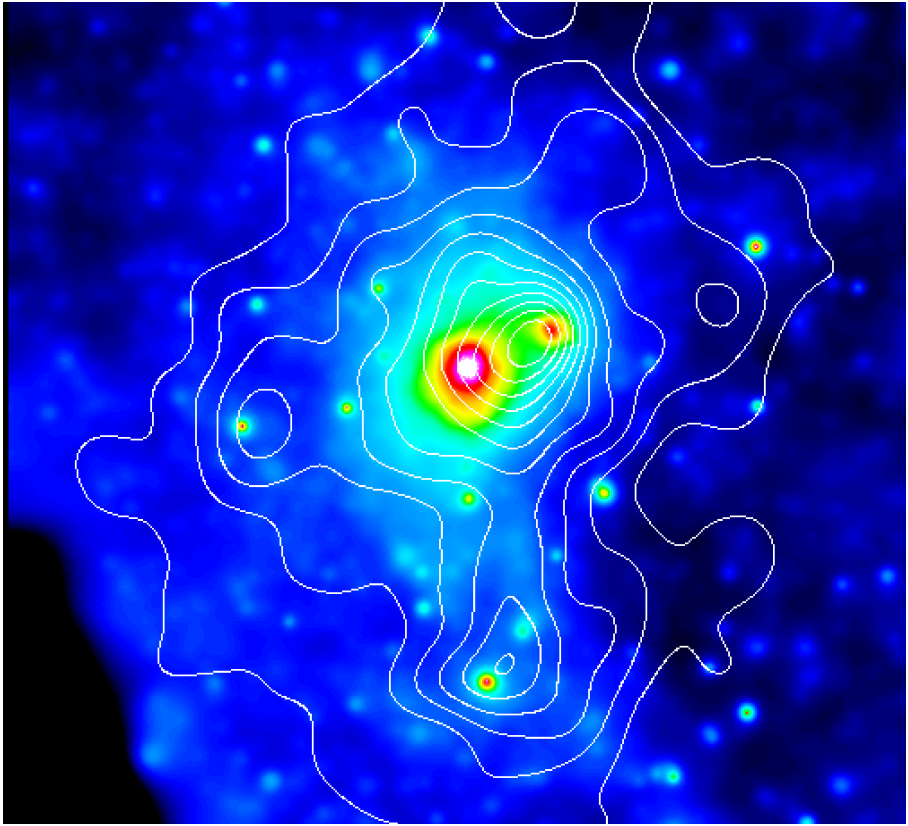,width=12cm,clip=} 
\vskip.2in
\caption{Comparison of the X-ray and the optical appearance of the
Virgo cluster. The colour image shows the X-ray emission as observed
by the ROSAT All-Sky Survey in the hard band (0.5-2.0 keV). It is smoothed with a Gaussian filter with
$\sigma = $24~arcmin on the faintest levels and decreasing filter size
with increasing surface brightness. The contours show the number
density of the 1292 
member galaxies of the VCC smoothed again with a Gaussian of
$\sigma = $24~arcmin. The spacing of the contours is linear. The
lowest contour line and the contour spacing is $1.5\times10^{-3}$
galaxies per arcmin$^{2}$. 
The image has a size of
12.8$^{\circ}\times$12.8$^{\circ}$. North is up and West to the right.
The main X-ray maximum is centred on
M87 ($\alpha_{2000} \approx$ 12$^{\rm h} 30^{\rm m}$, $\delta_{2000} \approx
12^{circ}$23~arcmin). For the identification of further galaxies, cf.~Fig.~2. 
}
\end{figure}

Figures 1 and 2 show the distribution of both components. The Virgo cluster
is obviously a complex system. There
is a pronounced double structure in the direction N-S. The Northern clump,
called ``cluster A'' in (Binggeli et al. 1987, BTS87), is dominated by
M87, which coincides with the maximum of the X-ray emission in the
whole area. The Southern, much less pronounced clump, coincides with
M49, another supergiant  
elliptical that is even slightly brighter than M87. However, as emphasized
in BTS87, the galaxy density contours do not peak on M87 but almost 
$1^{\circ}$ WNW of it -- more than halfway in the direction to M86, which is 
yet another giant E galaxy
(see Fig.~2a).
The reason for this ``mispointing'' of M87 is now clear: the Northern clump
itself is a double system comprising the dominating, massive subcluster
centred on M87 and a smaller subcluster centred on M86. This view is
supported by the existence of a swarm of low-velocity dwarf galaxies around
M86 (which itself has a negative velocity, Binggeli et al.~1993) and the 
extended X-ray halo of M86 (Forman et al. 2001). The M86 subcluster
seems to fall into the M87 subcluster from behind.

\begin{figure}[ht]
\psfig{figure=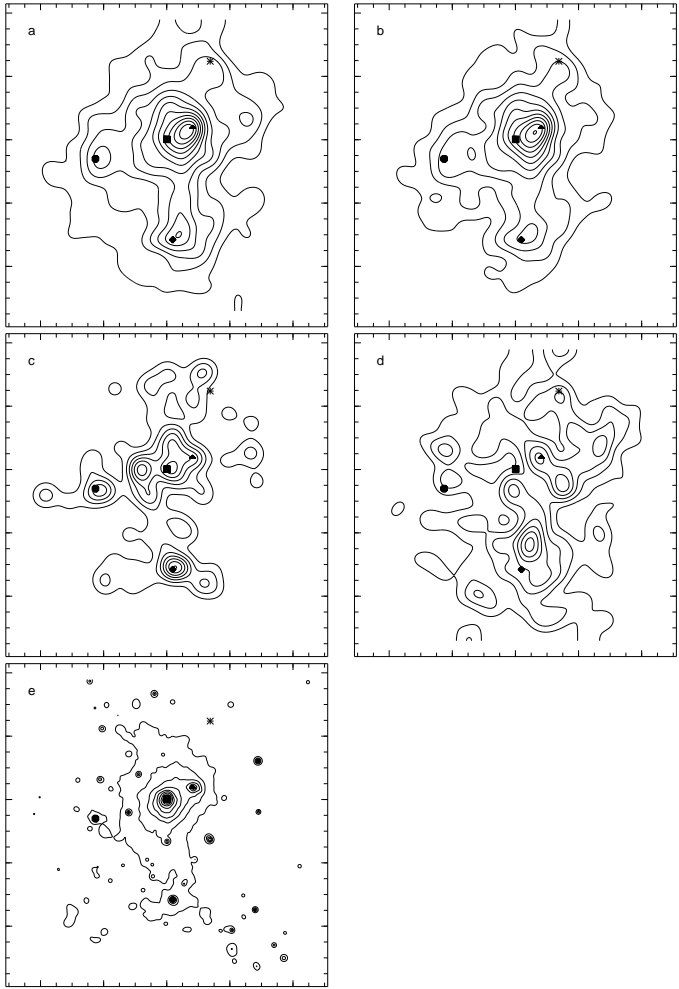,width=9cm,clip=} 
\vskip.2in
\caption{
Distributions of the different galaxy types. Number densities
smoothed with a Gaussian of $\sigma = $24~arcmin are shown. (a) all
galaxy types, (b) dwarf elliptical and dwarf S0 galaxies, (c)
elliptical and S0 galaxies, (d) spiral and irregular galaxies.
The contours are linear with spacings of (a) $1.5\times10^{-3}$
galaxies/arcmin$^{2}$, (b) $1.3\times10^{-3}$ galaxies/arcmin$^{2}$, 
(c) $2.2\times10^{-4}$ galaxies/arcmin$^{2}$, 
(d) $3.6\times10^{-4}$ galaxies/arcmin$^{2}$. 
The level of the first contour line is
equal to the spacing. For better
comparison with Fig.~1 the positions of five galaxies are marked: M87
(square), M49 (diamond), M86 (triangle), M60 (octagon), M100
(star). (a) is the same as the contours in Fig.~1. For comparison the
X-ray image is shown again in (e) with logarithmically spaced contours.
The size of each image is 12.8$^{\circ}\times$12.8$^{\circ}$. The
distance between two tick marks is 42~arcmin.
}
\end{figure}

We thus deal essentially with three major subclusters centred on the giant ellipticals M87, M86, and
M49. They are readily visible as maxima in Fig.~1. Note that these
X-ray sources are not point sources but are extended: the hot gas is located
in the potential of the (extended) subclusters. 

The number density distribution for the various galaxy types in the Virgo
cluster is shown in Fig.~2b (early-type dwarfs, i.e.~dE and dS0), 2c 
(early-type giants, i.e.~E and S0), and 2d (late types, i.e.~spirals and
irregulars). Confirming the results of BTS87
we find very different distributions for the different types.
The dwarf ellipticals (Fig.~2b) have a distribution very
similar to that of all galaxies (Fig.~2a), simply
because they make up 3/4 of the total cluster population. The
ellipticals and S0s (Fig.~2c) show a less extended distribution and are more
concentrated to the subcluster centres, i.e. the
X-ray maxima (compare with Fig.~2e). In sharp contrast, 
the distribution of the spirals and irregulars (Fig.~2d) is
very extended and shows no correlation with the X-ray emission.
Note also the difference in the type-mix between the various clumps.
The M49 subcluster is obviously spiral-rich.

For both components -- gas and galaxies -- we have fitted the observed subcluster 
profiles with isothermal $\beta$-models, which allow an easy deprojection to 
three-dimensional densities.
The comparison of the X-ray and optical density profiles of the Virgo 
subclusters has led to the following results:

-- The X-ray profile is steeper than the optical profile in the central part
($r <$ 70 arcmin) but with a slight inverse trend at larger radii -- 
for both the M87 and M49 subclusters.
This is reflected in the much smaller X-ray core radii and the smaller
X-ray $\beta$-values than the corresponding optical values. 

\begin{figure}[ht]
\psfig{figure=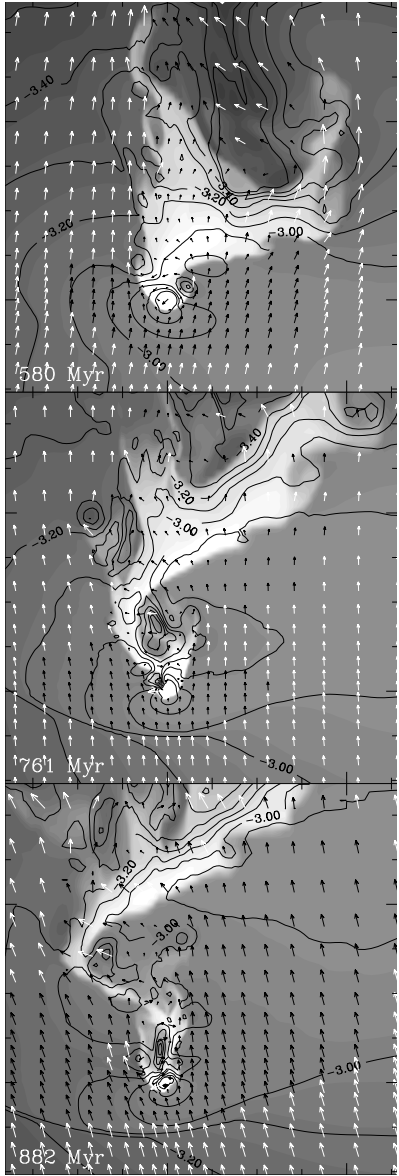,width=5.2cm,clip=} 
\vskip.2in
\caption{Gas density (grey scale) and pressure (contours) of a galaxy
moving downwards towards the cluster centre. The arrows show the Mach
vectors (white when $M>1$, black otherwise). The gas of the galaxy is
stripped due to ram pressure. 
}
\end{figure}

\begin{figure}[ht]
\psfig{figure=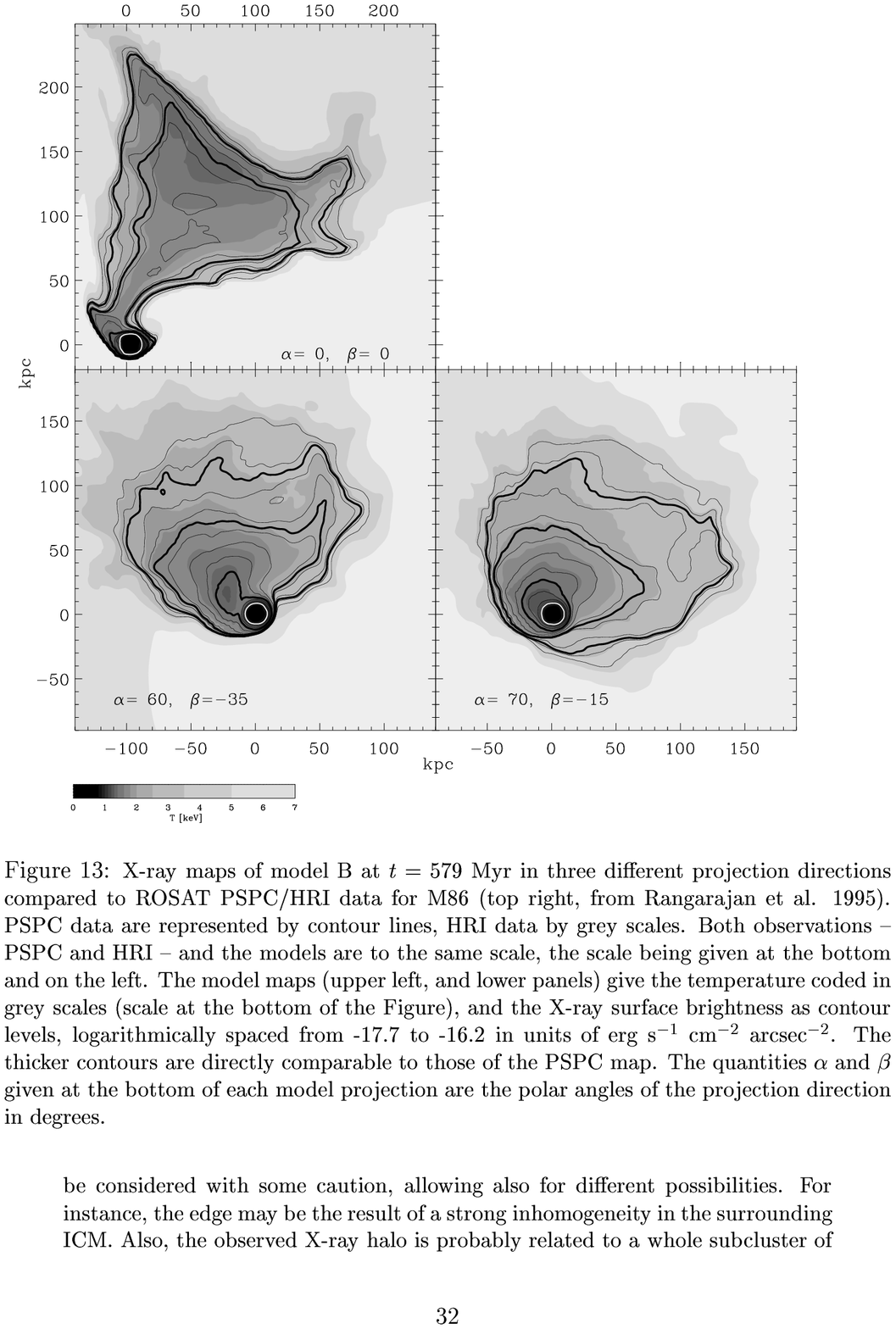,width=15cm,clip=} 
\vskip.2in
\caption{Simulated X-ray images (contours) and temperatures (grey
scale) of a galaxy affected by ram-pressure
stripping in three different projection directions.} 
\end{figure}

\begin{table}
\caption[]{Masses of the subclusters centred on M87, M49 and M86
(integrated values). Galaxy masses are based on a
constant $\msol/\lsol$ of 20. The total masses
of the subclusters were calculated on the assumption of an
isothermal gas. The 1$\sigma$ 
error for the galaxy masses is $\approx$ 20\%, that for the total masses
is 30-50\% (except for M86).
}
\begin{center}
\begin{tabular}{|c|c|c|c|c|c|}
\hline
&\multispan2\hfil M87 \hfil&\multispan2\hfil M49 \hfil& M86        \cr
& ($r$=400kpc) & ($r$=1Mpc) & ($r$=400kpc)&($r$=750kpc) & ($r$=240kpc) \cr
\hline
$M_{tot} [10^{13}\msol]$ & 5.5 & 14   & 4.7 & 8.7  &  $1-3$ \cr
$M_{gas} [10^{13}\msol]$ & 0.42& 1.9  &0.026&0.044 &  -     \cr
$M_{gal} [10^{13}\msol]$ &0.14& 0.51 & 0.17&0.34  & 0.06   \cr
$M_{gas}/M_{tot}$        & 8\% & 14\% &0.6\%&0.5\% &  -     \cr
$M_{gal}/M_{tot}$        & 3\% &  4\% & 4\% & 4\%  & $2-6$\%\cr
$M_{gal}/M_{gas}$        &0.34 & 0.28 & 6.6 & 7.7  &  -     \cr
$M/L [\msol/\lsol]$      &$\approx 500$ &$\approx 500$&$\approx 500$
&$\approx 600$  & $300-1000$\cr
\hline
\end{tabular}
\end{center}
\end{table}

-- The more massive the subcluster the less compact is its structure - both in
optical and in X-rays, i.e.~poorer subclusters have a steeper radial profile.
This behaviour is in agreement with the results
of N-body simulations by Navarro et al. (1997) who also found  
steeper profiles for
low-mass haloes than for high-mass haloes.
The same systematic effect is well-known to hold also for normal
elliptical galaxies, i.e. low-luminosity ellipticals
(like M32, but unlike ``dwarf'' ellipticals) appear much more compact
than giant ellipticals. 

-- Different Hubble types show different slopes in the subcluster
   profiles. We confirm the results by BTS87 that
   early-type galaxies are much more strongly clustered than late-type
   galaxies, reflecting of course the well-known 
   morphology-density relation (Dressler 1980).

-- There is a region South-West of M87 which shows a steeper gradient
   than the rest of the M87 subcluster - both in the optical and in
   X-rays. Such steep gradients in the X-ray emission can be caused by
   shock waves emerging after the collision and merging of subclusters
   (Schindler \& M\"uller 1993). 

-- Differential and integrated profiles of the galaxy mass, gas mass, 
   and total gravitating mass density are studied
   for both the M87 subcluster and the M49 subcluster (see Table 1). 
   The gas
   mass fraction in M87, with 8\% and 14\% at 400~kpc and 
   1~Mpc, respectively, is slightly on the low side for clusters, but
   is still in the normal range.
   One finds
   the usual behaviour that the gas distribution is somewhat flatter than the
   distribution of the total mass, i.e. the gas distribution is more
   extended. 
   The integrated gas mass is about 3 times the galaxy mass. 
   This is about the same factor that was found in other clusters.
   A comparison of the gravitating mass of the M87 subcluster
   ($M_{tot,M87}=2.1\times10^{14}\msol$) with the masses of other 
   clusters shows that they are in general more massive than the M87
   subcluster, e.g. the Coma and the Perseus cluster
   have almost a factor of ten more gravitating mass.
   The galaxy mass
   density is getting flatter towards the centre, which is the reason
   why the mass-to-light ratio also 
   tends to increase with decreasing radius (excluding M87 itself,
   i.e.~for $r <$ 60~arcmin). 
   We find
   projected mass-to-light ratios between 300 and 500 $\msol/\lsol$ at radii
   larger than 200~kpc which are relatively large mass-to-light ratios
   compared to the
   values of 90-250 $h_{50} \msol/\lso$ generally found in other clusters. 

   The mass distributions for the M49 subcluster are similar. 
   The only quantity which differs substantially between the M87 and M49 
   subclusters, however, is the
   gas mass fraction. For the M49 subcluster we find a very small gas
   mass fraction of less than 1\% of the total mass.

For more details on the comparison between the gas and the galaxy distribution
see Schindler et al. (1999).

\section{Ram-pressure stripping of M86-like galaxies}

We performed hydrodynamic simulations of ram-pressure stripping of
elliptical galaxies as they pass through the intra-cluster gas. We
considered different orbits of the galaxies through the cluster. 

An
example of such a stripping process is shown in Fig.~3. We find Kelvin-Helmholtz
wiggles developing on the ISM-ICM interface, which produce secondary
shocks and secondary rarefaction fans on the sides downstream of the
galaxy centre. Irregular extensions of the ISM form, which grow or
stretch until they break away. Initially, the momentum acquired by the
ISM is spent in climbing up the galaxy's potential well, so that for
some time the ISM is almost co-moving with the galaxy. As soon as the
ISM is displaced and decelerated, it starts falling towards the
cluster centre. A second shock is formed in front of the 
stripped, flowing material. The pressure built up behind the second
shock in  turn displaces the ISM upstream. The produces the
``S''-shape of the trail visible in Fig.~3.

We find 
that the gas cannot only be stripped off as the galaxy approaches the
cluster centre, but the galaxy can again accumulate some gas when it is
in the apocentre of its orbit. 

Also the  X-ray morphologies of
the stripped galaxies and their X-ray temperature maps can be calculated (Fig.~4) for direct comparison
with observations, like e.g. the X-ray morphology and temperature map of M86 as observed
by CHANDRA (Forman et al. 2001)

For more details on the hydrodynamic simulations of the ram-pressure
stripping process see Toniazzo \&
Schindler (2001).

\bigskip\noindent
{\bf References}

\noindent
\refs

Binggeli B., Tammann G.A., Sandage A., 1987, AJ 94, 251 (= BTS87)

Binggeli B., Popescu C.C., Tammann G.A., 1993, A\&AS 98, 275

Dressler A., 1980, ApJ 236, 351

Forman W., et al., 2001, {\it 
Proceedings of the XXI Moriond Conference: 
Galaxy Clusters and the High Redshift Universe Observed in X-rays},
Neumann D.M. (ed.)

Navarro J.F., Frenk C.S., White S.D.M., 1997, ApJ 490, 493

Schindler S., Binggeli B., B\"ohringer H., 1999, A\&A 343, 420

Schindler S., M\"uller E., 1993, A\&A 272, 137

Toniazzo T., Schindler S., 2001, MNRAS 325, 509

\endrefs

\end{document}